\documentclass{aa}

\usepackage{comment}
\usepackage{graphicx}
\usepackage{natbib}
\usepackage{float}
\usepackage{txfonts}
\usepackage[switch]{lineno}

\bibpunct{(}{)}{;}{a}{}{,} 
\usepackage{xcolor}  

\usepackage{hyperref}
\hypersetup{
  colorlinks=true,  
  linkcolor=blue,   
  citecolor=blue,  
  urlcolor=blue,    
}
\usepackage{bm}  
\begin{document} 

   \title{A systematic search for spectral hardening in blazar flares with the {\it Fermi}-Large Area Telescope}

   \author{A. Dinesh,
          \inst{1}
          A. Dominguez,\inst{1}
          V. Paliya, \inst{2}
          J. L. Contreras, \inst{1}
          S. Buson, \inst{3,4}
          M. Ajello \inst{5}
          }

   \institute{IPARCOS and Department of EMFTEL, Universidad Complutense de Madrid, E-28040 Madrid, Spain
              \email{adinesh@ucm.es, alberto.d@ucm.es}
         \and
             Inter-University Centre for Astronomy and Astrophysics (IUCAA), SPPU Campus, 411007, Pune, India
        \and 
             Deutsches Elektronen-Synchrotron DESY, Platanenallee 6, 15738 Zeuthen, Germany
        \and
             Julius-Maximilians-Universit\"at W\"urzburg, Fakultät f\"ur Physik und Astronomie, Institut f\"ur Theoretische Physik und Astrophysik, Lehrstuhl f\"ur Astronomie,  Emil-Fischer-Str. 31, D-97074 W\"urzburg, Germany
        \and
             Department of Physics and Astronomy, Clemson University, Kinard Lab of Physics, Clemson, SC 29634-0978, USA
             }

   \date{Received September 15, 1996; accepted March 16, 1997}

   \date{Received ; accepted }

\authorrunning{Dinesh et al.}
\titlerunning{Systematic search for spectral hardening}

\abstract
{Blazars, a subclass of active galactic nuclei (AGNs), are among the most powerful and variable $\gamma$-ray sources in the universe. They emit non-thermal radiation across the electromagnetic spectrum in the form of relativistic jets, characterized by rapid flux and polarization variability. High synchrotron-peaked blazars (HSPs) and extreme high synchrotron-peaked blazars (EHSPs), with synchrotron peaks exceeding $10^{15}$ Hz and $10^{17}$ Hz, respectively, are crucial for understanding the full range of blazar phenomena and testing models of jet physics. Yet, their understanding remains challenging.}
{This work aims to systematically identify and characterize the most extreme $\gamma$-ray blazars using data from the Large Area Telescope (LAT) on board the \textit{Fermi Gamma-ray Space Telescope}. The focus is on spectral hardening, where the $\gamma$-ray spectrum becomes harder at higher energies, particularly during flaring episodes. This represents the first dedicated analysis of spectral hardening, as previous studies have only explored this phenomenon in a few individual sources.}
{We analyzed a sample of 138 blazars selected from the 4FGL-DR2 catalog with high synchrotron peak frequencies and well-sampled light curves. Flaring periods were selected using Bayesian Block analysis. Each flare was then analyzed through $\gamma$-ray spectral fitting with both power-law and broken power-law models to identify potential spectral hardening. The significance of spectral hardening was assessed using a test statistic, TS$_{\mathrm{hardening}}$, based on the likelihood ratio of the two spectral models.}
{We identified two flaring episodes with indications of spectral hardening, one in 4FGL J0238.4$-$3116 and another in PKS 2155$-$304, the latter detected independently by both selection methods but referring to the same flaring period. This number of candidate events is consistent with expectations from statistical fluctuations, suggesting that spectral hardening is, at most, a rare occurrence in $\gamma$-ray blazars. These results provide the first population-level constraint on the frequency of such events (< 0.1\%). The scarcity of events reinforces the notion that the dominant blazar emission mechanism is well described by smoothly varying power-law spectra across the {\it Fermi}-LAT range, with sharp spectral hardenings representing rare deviations likely tied to exceptional jet conditions or transient physical processes. Although these flares show notable spectral changes, their statistical significance remains modest and motivates future multi-wavelength studies to assess whether these rare flares reflect genuinely distinct physical processes within blazar jets.
}
{}


\maketitle
\section{Introduction}

Blazars, a subclass of active galactic nuclei (AGNs) with relativistic jets aligned close to our line of sight, are among the most powerful and variable sources of $\gamma$-rays in the universe \citep[e.g.,][]{urry1995unified}. These jets emit non-thermal radiation across the entire electromagnetic spectrum, from radio to $\gamma$-rays, driven by relativistic particles within the jet \citep[e.g.,][]{ulrich1997variability}. Blazars are characterized by rapid flux variability, high polarization, and apparent superluminal motion \citep[e.g.,][]{jorstad2001multiepoch, marscher2008core}.

The broadband spectral energy distribution (SED) of blazars typically exhibits a double-peaked structure in the $\nu L_{\nu}$ vs. $\nu$ plane (i.e., power emitted per unit logarithmic frequency interval versus frequency). The lower peak is associated with synchrotron radiation from relativistic electrons and positrons in the jet's magnetic field, while the origin of the higher-energy peak remains debated. Leptonic models attribute this peak to inverse Compton (IC) scattering, either within the jet, known as synchrotron self-Compton \citep[SSC; e.g.,][]{ghisellini1989bulk, maraschi1992jet, 2010MNRAS.401.1570T, paliya2018leptonic, paliya2019detection, van2019systematic}, or involving external photon fields, known as external Compton \citep[EC; e.g.,][]{2009ApJ...704...38S, lefa2011formation, lewis2019electron}. In contrast, hadronic models propose that $\gamma$-rays are produced by ultrarelativistic protons through interactions that generate secondary particles which decay into $\gamma$-rays \citep[e.g.,][]{1992A&A...253L..21M, bottcher2013leptonic, 2017A&A...602A..25Z, 2022MNRAS.509.2102G}.

Blazars are classified into BL Lac objects (BL Lacs) and flat-spectrum radio quasars (FSRQs) based on their emission line properties, with weak or absent emission lines in the former and strong lines in the latter in their optical spectra \citep[e.g.,][]{urry1995unified, padovani2017active}. There is also an additional sub-classification based on the synchrotron peak frequency \citep[e.g.,][]{abdo2010spectralb}. High synchrotron-peaked blazars (HSPs) and extreme high synchrotron-peaked blazars (EHSPs) are particularly noteworthy, with synchrotron peak frequencies exceeding $10^{15}$ Hz and $10^{17}$ Hz, respectively \citep[e.g.,][]{costamante2001extreme, 2019ApJ...882L...3P}. These blazars are critical for understanding the full range of blazar phenomena and for testing models of jet physics, particularly in the context of the blazar sequence, which attempts to order blazars by luminosity and synchrotron peak frequency \citep[e.g.,][]{fossati1998unifying, ghisellini2008blazar, ghisellini2017fermi}. However, the blazar sequence has been questioned, with suggestions that observational biases may influence the perceived trends \citep[e.g.,][]{giommi2012simplified, chang20193hsp, 10.1093/mnras/stab1182}. The blazar sequence framework is helpful for understanding the diverse properties of blazars and how they may be related to the physical conditions within their relativistic jets. However, a comprehensive understanding of HSP and EHSP blazars remains challenging due to their intrinsic diversity and limited availability of sources with detailed and simultaneous multi-wavelength spectral coverage \citep[e.g.,][]{10.1093/mnras/sty857,foffano19,costamante20,nievas2022hunting,2024A&A...685A.117M, 2502.07940v1, 2025ApJ...988L..50B, 2025A&A...700A.229L}. Observing these sources is essential for improving our understanding of the blazar sequence and blazar evolution. Note that a blazar may show the properties of an EHSP temporarily during flaring states, as these events can shift the synchrotron peak toward higher energies. Therefore, identifying such transient states is crucial for a complete characterization of EHSPs \citep{valverde20, abe2024insights}.

Furthermore, according to \cite{biteau2020progress} and \cite{galaxies10010035}, a majority of known TeV blazars show extreme synchrotron peaks. Therefore, they conclude that TeV blazars may have a greater tendency to display extreme behaviors in terms of SED peaks. As very high energy (VHE, $E>50$ GeV) extragalactic particle accelerators, extreme blazars represent prime candidates for exploring multi-messenger correlations with neutrinos and cosmic rays \citep[e.g.,][]{costamante2001extreme}. Given their hard spectrum at higher energies, these blazars are important for studying cosmic $\gamma$-ray photon propagation, including the extragalactic background light and cosmological research \citep[e.g.,][]{dominguez13, dominguez15, dominguez19, saldana-lopez21, dominguez24}. They may also be the dominant component of the extragalactic TeV background \citep[e.g.,][]{giommi2015simplified, ajello2015origin}. All this makes understanding EHSPs a fundamental topic and desirable targets for current and future VHE telescopes such as the Cherenkov Telescope Array Observatory \citep[CTAO;][]{acharya2017science}. In summary, despite numerous individual studies, a complete and uniform characterization of extreme blazars across a large sample remains limited, highlighting the value of systematic searches and population-level analyses \citep[e.g.,][]{biteau2020progress, galaxies10010035}.

In this study, we used \textit{Fermi}-LAT data to search for spectral hardening in a large sample of HSP blazars. Similar behavior has been reported in radio galaxies \citep{abdo10_magn,brown17,rulten20}, in the HSP 1ES 0502+675 \citep{zeng2022spectral}, and in an FSRQ \citep{paliya2025detection}, but it has not been systematically explored in HSP blazars, where synchrotron peaks may shift into the LAT band during flares. We employed Bayesian Block Analysis to identify robust flaring periods and test for the presence of spectral hardening by comparing fits of power-law versus broken power-law models to the gamma-ray spectra. Such spectral hardening could hint at complex conditions within blazar jets, such as transitions between different emission regions or electron populations, and deserves detailed multi-wavelength investigations \citep[e.g.,][]{abramowski2012multiwavelength, 10.1093/mnras/stab1445}. Previous studies have reported similar features, but their occurrence is rare and often linked to specific flaring events \citep{abdo10_magn, zeng2022spectral}. By systematically searching for this feature in a large sample of blazars, we aim to enhance our understanding of the most extreme blazar behaviors and their underlying physical mechanisms.

The structure of the paper is as follows: Section \ref{sec2} outlines the source selection criteria and the analysis of \textit{Fermi}-LAT data. In Section \ref{sec3}, we detail the methodology, including the flare selection process and the search for the spectral hardening feature. Section \ref{sec4} presents the analysis and discussion of our results. Finally, Section \ref{sec5} provides a summary and conclusions.

\section{Data analysis}\label{sec2}

\subsection{Source selection}
\label{sec2.1} 
In this section, we describe the criteria used for selecting our sample of blazars. The sources were drawn from the 4FGL-DR2 catalog \citep{ballet2020fermi}, for which light curves (LCs) have been previously computed and utilized by \citet{penil2025search}, \citet{penil24a}, \citet{penil24b}, \citet{rico25}, and \citet{penil25}. These LCs include all AGN types (3308 AGNs) and cover the first 12 years of \textit{Fermi}-LAT data in 28-day binning. We applied the following selection criteria: (1) High-latitude \textit{Fermi}-LAT sources $(|b|>10^\circ)$, to minimize contamination from diffuse emission in the Galactic plane. (2) A synchrotron peak frequency greater than $10^{16}$ Hz,  based on synchrotron peak frequencies taken from \citet{yang2022spectral}, and from the 4LAC–DR3 catalog \citep{ajello2022fourth} only when no value is available in the former. The reason is that \citet{yang2022spectral} developed an analysis specifically focused on synchrotron peaks, though with possible uncertainties due to non-simultaneous data, which also affect the 4LAC estimates (see references for details). Using these two criteria, we obtain 365 sources. Finally, we require that the LCs contain less than 50\% upper-limit (UL) bins, meaning that more than half of the time bins must be detections\footnote{We define a time bin as a detection if its Test Statistic (TS) exceeds 1, and otherwise consider it an UL. See Sec.\ref{sec3.1} for further discussion.} to allow for reliable flare identification \citep[e.g.,~][]{penil20}. We note explicitly that this condition of requiring fewer than 50\% upper limits per flare biases our analysis toward brighter flares and more luminous sources. As a consequence, our results primarily characterize variability and spectral features in relatively bright events, and caution should be exercised when generalizing these findings to intrinsically faint HSP and EHSP sources. This results in a final sample of 138 blazars, with synchrotron peak frequencies taken from \citet{yang2022spectral} for 129 sources and from 4LAC-DR3 for 9 sources. Of these 138 blazars, 20 have synchrotron peak frequencies greater than $10^{17}$ Hz.

 The selection of sources with synchrotron peak frequencies above $10^{16}$ Hz is motivated by our aim to identify blazars that may exhibit spectral hardening in the $\gamma$-ray band. Blazars with higher synchrotron peak frequencies are more likely to show spectral hardening within the energy range detectable by \textit{Fermi}-LAT, as their synchrotron emission extends further into higher energies, increasing the probability of encountering a spectral transition in the framework of synchrotron/SSC models \citep{ghisellini2017fermi}. By selecting blazars with already high synchrotron peak frequencies, we also reduce the need for large shifts during flaring states to observe such spectral hardening within the \textit{Fermi}-LAT band. This threshold ($\nu_{\text{peak}} > 10^{16}$ Hz) includes both classical HSP and EHSP blazars. The rationale is to investigate whether spectral hardening observed during flaring episodes could occasionally correspond to synchrotron peak frequencies temporarily moving above $10^{17}$ Hz, thus producing transient EHSP-like states \citep[e.g.,][]{valverde20,abe2024insights}. We emphasize that synchrotron peak values are used only for the initial source selection, while our $\gamma$-ray spectral analysis and identification of spectral hardening are carried out solely with \textit{Fermi}-LAT data, independent of assumptions about synchrotron peak positions.

\subsection{Gamma-ray data reduction and spectral fitting} \label{sec2.3}
We performed the analysis of the $\gamma$-ray data using the \texttt{Fermipy} software package (version 1.3.1), an open-source Python package designed to analyze \textit{Fermi}-LAT data \citep{wood2017Fermipy}. The analysis followed the criteria for a point source analysis, utilizing the latest Galactic and isotropic diffuse models, {\it gll\_iem\_v07.fits} and {\it iso\_P8R3\_SOURCE\_V3\_v1.txt}, respectively\footnote{\href{https://Fermi.gsfc.nasa.gov/ssc/data/access/lat/BackgroundModels.html}{https://Fermi.gsfc.nasa.gov/ssc/data/access/lat/BackgroundModels.html}}. The analysis is done using the latest version of the \textit{Fermi} Science Tools (V2.2.0) and the instrument response function used is P8R3 SOURCE V3 \citep{2013arXiv1303.3514A}.

We conducted the spectral fitting over flaring periods, which are identified as described in Section \ref{sec3.1}. For each source in our sample, we defined a region of interest (ROI) with a radius of 15 degrees centered on the source. The filter expression used is `(DATA QUAL>0)$\&\&$(LAT CONFIG == 1)$\&\&$(angsep(RA source, DEC source, RA\_SUN, DEC\_SUN)>15)', where RA source and DEC source are the right ascension and declination of the source, and RA\_SUN and DEC\_SUN are those of the Sun. This selection ensures that only good time intervals (GTIs) are considered, and time periods when the source is within 15 degrees of the Sun are excluded to prevent contamination from solar emission. 

The $\gamma$-ray sky model was constructed using the 4FGL-DR3 catalog, and the energy range considered was 50 MeV to 300 GeV. We used different PSF event types and zenith angle cuts for different energy ranges to optimize angular resolution and sensitivity:

\begin{itemize}
    \item 50 MeV to 100 MeV: PSF3 event types with zenith angles less than 80$^\circ$.
    \item 100 MeV to 300 MeV: PSF2 and PSF3 event types with zenith angles less than 90$^\circ$.
    \item 300 MeV to 1 GeV: PSF1, PSF2, and PSF3 event types with zenith angles less than 100$^\circ$.
    \item 1 GeV to 300 GeV: PSF0, PSF1, PSF2, and PSF3 event types with zenith angles less than 105$^\circ$.
\end{itemize}

This strategy provides the best possible angular resolution and minimizes background contamination at low energies, ensuring that our gamma-ray data analysis is as robust as possible. Also, this selection maximizes sensitivity to point sources like blazars, which is crucial for accurate spectral and spatial analyses. 

The normalization of the Galactic and isotropic diffuse component was left free. Additionally, the normalization parameter of catalog sources with a test statistic (TS) value greater than 25 within 9 degrees of the ROI center is set free. For sources with TS $>$ 500 within 12 degrees, both the normalization and spectral index are allowed to vary, accounting for long-term averaged source properties over the first 12 years of {\it Fermi}-LAT data. After an initial fitting, a source-finding step is implemented using gta.find\_sources from \texttt{Fermipy} to identify additional unresolved sources in the ROI. Following this, the same TS-based criteria are reapplied to both catalog and newly identified sources during the analysis period, refining the model to account for temporal variability or new source activity. Finally, a global fit is performed to obtain the optimized model parameters.

We then tested the broken power-law (BPL) model as a candidate spectral function. Initially, we fitted a power-law (PL) spectral function and recorded the likelihood of the fit, $\mathcal{L}_{PL}$. Subsequently, we changed the spectral model to a broken power-law(BPL) function, defined as

\begin{equation}\label{eq1}
\frac{dN}{dE} = N_0 \times \left\{ \begin{array}{ll} (E/E_{\rm break})^{-\Gamma_{1}} & \mbox{if $E < E_{\rm break}$} \\ (E/E_{\rm break})^{-\Gamma_{2}} & \mbox{otherwise} \end{array} \right. ,
\end{equation}

\noindent where $E_{\rm break}$ is the break energy (in MeV), $\Gamma_{1}$ is the photon index for $E \leq E_{\rm break}$, and $\Gamma_{2}$ is the photon index for $E \geq E_{\rm break}$. We iterated the fit by fixing the break energy, $E_{\rm break}$, at 100 logarithmic intervals between 0.1 GeV and 10 GeV, calculating the likelihood, $\mathcal{L}_{BPL}$, at each step. The photon indices ($\Gamma_{1}$ and $\Gamma_{2}$) are allowed to vary between -5 and 5. At each iteration, we calculated the value of TS$_{\mathrm{hardening}}$ = 2 $\times$ ($\mathcal{L}_{BPL}$ - $\mathcal{L}_{PL}$), which quantifies the test statistics of spectral hardening from comparing the BPL model likelihood with the PL model likelihood. Note that this is similar, but more specific, to the TS$_{\mathrm{curv}}$ used in the 4FGL catalog for determining whether a curved spectral function is preferred over a PL function \citep{abdollahi2020fermi}. And while creating the SEDs, we left the parameters of all sources free within 3 degrees. We finally selected the $E_{\rm break}$ that maximized the TS$_{\mathrm{hardening}}$ profile.

The choice of a BPL function is both statistically and physically motivated. Flaring periods are characterized by rapid spectral changes and generally short integration times, making smoother functions such as a logparabola less effective for constraining spectral hardening \citep[e.g.,~][]{abdo2010spectralb,ackermann11}. Furthermore, a BPL, with an abrupt spectral hardening and two independent photon indices, better captures sudden changes in particle acceleration or cooling while providing a clearer quantification of spectral hardening \citep[e.g.,~][]{tavecchio2009hard}.

In summary, we defined spectral hardening as cases where the {\it Fermi}-LAT spectrum (50 MeV -- 300 GeV) is significantly better fit by a broken power law than by a single power law, with the photon index above the break energy harder than below. This operational definition, consistent with previous {\it Fermi}-LAT studies \citep{abdo10_magn,brown17,rulten20,zeng2022spectral,paliya2025detection}, is applied here as a statistical method to identify localized slope changes, without invoking a specific physical interpretation.

\section{Methodology}\label{sec3}

We investigated the presence of spectral hardening during flaring episodes. These flaring episodes, characterized by increased activity, are selected from the LCs and are expected to enhance the probability of detecting the spectral hardening feature. The reason is that synchrotron peaks may shift toward higher energies during flares, and blazar spectra become harder in the $\gamma$-ray band when these sources are flaring \citep[e.g.,][]{2015ApJ...808..110A,2018A&A...620A.181A,valverde20,2020A&A...638A..14M,2021ApJ...906...91S,abe2024insights}. Secondly, our data analysis pipeline is run over the flaring periods, leading to a TS$_{\mathrm{hardening}}$ profile, which we used for evaluating the significance and robustness of the spectral hardening feature.

\subsection{Flare selection}\label{sec3.1}

To select flares in the LCs, we employed a systematic approach developed by \cite{2022icrc.confE.868W}. This method characterizes and models the variability of the LC using Bayesian Block Analysis\footnote{\url{https://docs.astropy.org/en/stable/api/astropy.stats.bayesian_blocks.html}} (BBA) as described by \cite{scargle2013studies}. BBA is a non-physical model designed to detect statistically significant variations in the data while minimizing the influence of observational or random uncertainties. Figure \ref{fig1} provides a visual representation of this procedure.

BBA divides the data into blocks by considering their variations, with the granularity of these blocks controlled by parameters such as $\gamma$ \citep{scargle2013studies}. For our analysis, we adopted $\gamma = 0.8$, which reflects a standard balance between detecting true variability in blazar LCs and minimizing false detections\footnote{\url{https://github.com/swagner-astro/lightcurves?tab=readme-ov-file}}. The number of detected flares depends on both the choice of $\gamma$ and the number of time bins in the LC, as these parameters influence the segmentation of the BBA algorithm. The corresponding $p_0$ value, which is another relevant parameter in BBA, is not fixed but is dynamically determined within the code, following the formulation by \citet{scargle2013studies}. This ensures that the block segmentation remains adaptive to the characteristics of each individual LC, and while variations in $p_0$ may lead to different flare counts, the adopted selection criteria ensure the robustness of our results. We only considered for the BBA, fluxes in time bins of the LC with TS > 1, which roughly corresponds to a $\sim1\sigma$ detection significance\footnote{For reference, TS values of 4, 9, and 25 correspond approximately to 2$\sigma$, 3$\sigma$, and 5$\sigma$ significance levels, respectively, assuming a $\chi^2$ distribution with one degree of freedom.}. While this threshold ensures a basic level of statistical significance in the LCs, we acknowledge that this criterion does not necessarily eliminate all outliers \citep{abdollahi2023fermi}. A more refined approach, such as filtering based on the $(F/\sigma_F)^2$ vs. TS relation, could improve the robustness of the selection process and represents a possible refinement for future analyses. However, since the most significant flaring periods tend to have high TS values, the difference between both approaches is not expected to substantially alter the overall results. All settings follow those described by \cite{2022icrc.confE.868W}.

We used two approaches to define a flare: the Baseline and Flip methods. Both are based on the HOP algorithm \citep{eisenstein1998hop}, which selects peaks as the highest local maxima of a flare, with the flare itself referred to as a HOP group. Bayesian blocks between two flares are defined as valley blocks. Throughout this work, we use the term \emph{flare} to denote periods of increased $\gamma$-ray activity, which in our analysis typically span from several months to over a year, reflecting the near-monthly binning of the LCs and the long-term variability patterns in blazars. Finally, to ensure the reliability of the flare identification, we require that each selected flare contains fewer than 50\% UL bins. This condition guarantees that the majority of the time bins are actual detections, which is essential for characterizing the flare structure and enabling robust spectral analysis. As a result of this filtering, the final sample includes only well-defined flares with reliable temporal profiles, suitable for a systematic spectral study.

\subsubsection{Baseline method}

The Baseline method, introduced by \cite{meyer2019characterizing}, defines a constant flux level as the baseline to select the onset of a flare. If the flux exceeds this baseline, the time period is categorized as a flare. Only blocks with flux values above the baseline are recognized as flares, and adjacent blocks surrounding a peak block that remain above the baseline are considered part of the same flare event within the limitations of the binning chosen. We used the mean flux of each LC as the baseline, consistent with the approach in \cite{2022icrc.confE.868W}. While this choice may result in fewer selected flares compared to using the quiescent or low-state flux, it provides a conservative approach that minimizes the inclusion of spurious events. Figure \ref{fig1} visually demonstrates the implementation of the Baseline method. While this method relies on the assumption of a baseline, it is a straightforward and effective approach for flare selection. This method identifies a total of 2193 flares; after applying the 50\% UL within-the-flare criterion, 1660 flares remain.

\subsubsection{Flip method}

\begin{figure}
\includegraphics[width=0.5\textwidth]{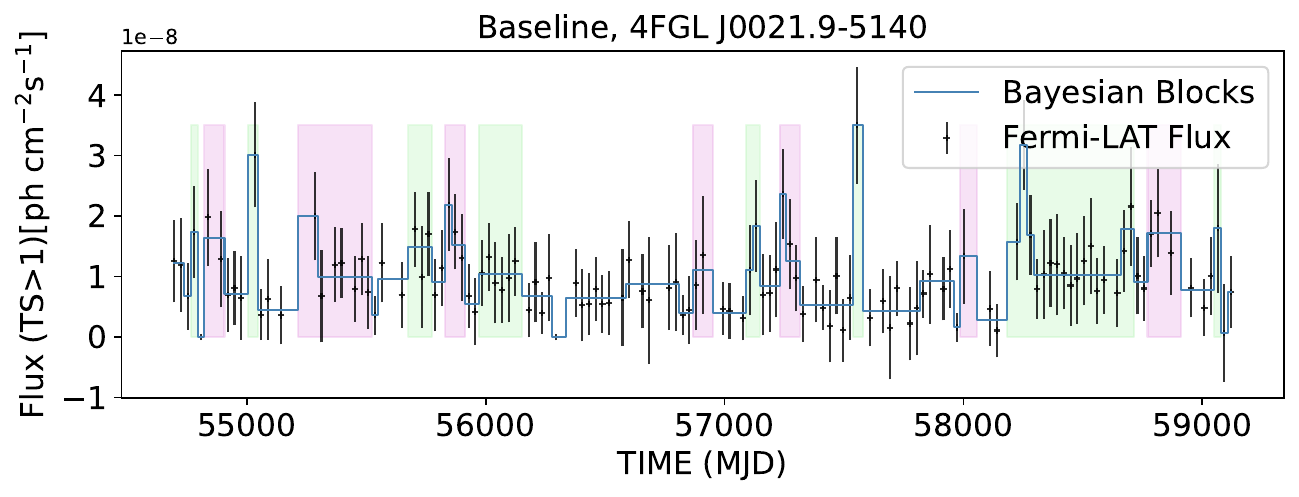}
\includegraphics[width=0.5\textwidth]{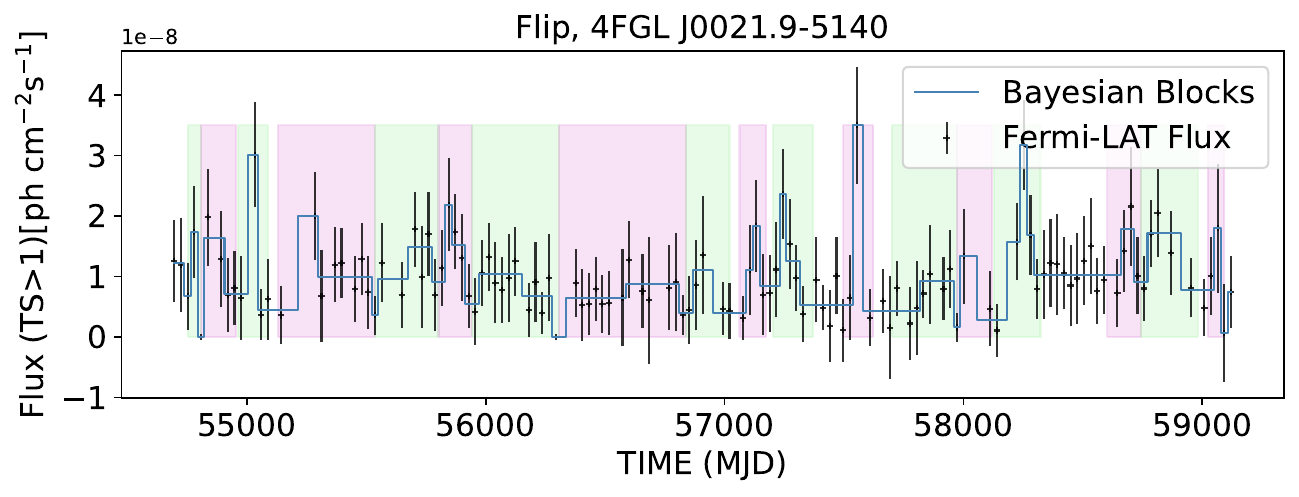}

\caption{Light curve of the blazar 4FGL J0021.9$-$5140, as an example, from \textit{Fermi}-LAT data (black flux measurements with error bars) with flare detection methods using the Baseline (top panel) and Flip (bottom panel) methods. We show the Bayesian blocks (blue line) and flares (shaded boxes, with different colors only for visualization convenience).}\label{fig1}
\end{figure}

The Flip method mitigates the bias associated with selecting a baseline by instead relying on the slope of data variation to determine the start and end of a flare. This approach extends the block adjacent to a valley block to include part of the valley block in the flare, with the remaining portion representing a quiescent or non-flaring level. If the adjacent block exceeds more than half of the valley block's length, only half of the valley block is included in the flare. This adjustment prevents overlap between adjacent flares. Figure \ref{fig1} shows an example of flare detection using the Flip method. A total of 2642 flares are initially identified using this method; applying the 50\% UL within-the-flare filter reduces the sample to 2403 flares.

Note that the Baseline method identifies flares based on a constant flux threshold, meaning only those exceeding a predefined level are considered. This approach can be conservative and may overlook shorter or modest flux increases. In contrast, the Flip method defines flare boundaries based on flux variations, making it more sensitive to transient activity that does not necessarily exceed a static threshold. As a result, it generally selects more flares, including those missed by the Baseline method, particularly when the defining feature is a change in slope rather than absolute flux level.

An additional observation is that, after applying the 50\% UL criterion, the Flip method retains about 90\% of the initially identified flares, compared to 75\% for the Baseline method. This higher retention rate suggests that the more flexible flare boundaries of the Flip method are better matched to periods with significant detections, providing a more robust basis for subsequent spectral analysis.

\subsection{Spectral hardening analysis framework}

The 28-day LCs used for flare selection are constructed using photons with energies above 100 MeV, as described in \citet{penil2025search}, and assuming a fixed photon index derived from a fit to the full 12-year dataset of each source. This index remains constant during the likelihood analysis in each time bin, ensuring that flare identification is based solely on flux variability and not influenced by short-term spectral changes. This approach avoids potential biases that could arise if the photon index were allowed to vary freely, which might skew the flare selection toward soft-spectrum intervals due to statistical noise in low-count bins. Although the use of a >100 MeV threshold could, in principle, introduce a minor selection bias when performing spectral fits over a broader energy range (50 MeV–300 GeV), this concern is mitigated by analyzing spectra over the full duration of each flare rather than on individual time bins. This strategy of integrating over the full flare duration helps mitigate the impact of statistical fluctuations in the LCs.  Moreover, the spectral hardening we detect corresponds to a distinct and localized change in slope, rather than a gradual curvature that might be induced by such a selection effect. Given the use of well-defined flares and the broad spectral fitting range, we consider any influence from the LC construction to be negligible for the objectives of this work.

After running our data analysis pipeline over all flaring periods selected by the Baseline and Flip methods, this is, a total of 4063 flares, we considered as candidate events those with TS$_{\mathrm{hardening}} \geq 12$. Since the BPL model introduces two additional degrees of freedom, TS$_{\mathrm{hardening}}$ is expected to follow a $\chi^2$ distribution with two degrees of freedom under the null hypothesis. We verified this assumption by comparing the empirical distribution of TS$_{\mathrm{hardening}}$ with the theoretical expectation and found good agreement. Given our sample of 4063 flares and the chosen threshold of TS$_{\mathrm{hardening}} \geq 12$, we expect approximately 10 false positives under the null hypothesis, based on the cumulative probability of the $\chi^2$ distribution with two degrees of freedom. For higher thresholds, such as TS$_{\mathrm{hardening}} > 16$, the expected number of false positives across the entire sample drops below one, providing a useful benchmark for identifying exceptionally rare events.  In this context, a TS$_{\mathrm{hardening}}$ value of 12 corresponds approximately to a 3$\sigma$ confidence level, and TS$_{\mathrm{hardening}} > 16$ to beyond 3.5$\sigma$, offering a familiar interpretation of the strength of the spectral deviations.

The three events with TS$_{\mathrm{hardening}} \geq 12$ undergo meticulous individual examination to complement the automatic safety checks that are applied to all flares, in order to: (1) manually verify the convergence of the fits, since rare cases of short integration times or limited statistics can still cause convergence issues even with the pipeline’s internal checks; (2) inspect the TS$_{\mathrm{hardening}}$ profile as a function of break energy, which should be smooth with a clear maximum; and (3) check for nearby LAT sources that may contaminate the spectrum. We analyzed whether $\gamma$-ray sources within the ROI could affect the result by examining whether any sources other than the target show varying parameters during spectral fitting. As an additional test, not part of the main analysis pipeline, we rerun the pipeline with these nearby sources fixed to assess whether the detection of spectral hardening is robust to potential contamination. This check is used to confirm that the result is not driven by variability in neighboring sources, but the default analysis keeps such sources free to ensure accurate error estimation. All these targeted checks return positive results, confirming that the identified spectral hardening features are robust and not artifacts of the fitting process or source confusion.

We tested the stability of the spectral indices for target blazars that have a bright nearby source (TS > 100) flaring simultaneously, focusing on cases where the target source's flare lasts longer than that of the nearby source. In these cases, the spectral indices remain consistent when comparing the time range in which only the target source is flaring to the full duration of the flare.

\section{Results and discussion}\label{sec4}
\subsection{Spectral hardening in blazar flares}

We identified three flares with indications of spectral hardening, all of which fall near or just above our detection threshold (see Table~\ref{tab1}). One flare is associated with 4FGL J0238.4$-$3116 (1RXS J023832.6$-$311658) with TS$_{\mathrm{hardening}}=12.0$, and two with 4FGL J2158.8$-$3013 (PKS 2155$-$304), detected independently by the Flip (TS$_{\mathrm{hardening}}=16.0$) and Baseline (TS$_{\mathrm{hardening}}=12.0$) methods. The flares in PKS 2155$-$304 overlap in time and likely correspond to the same physical event, captured slightly differently due to the distinct selection criteria of the two methods.

Given the small number of detections and their proximity to the significance threshold, these events are statistically compatible with fluctuations and do not provide strong evidence for the presence of spectral hardening across the blazar population. The flare in 4FGL J0238.4$-$3116 spans 231 days, while the event in PKS 2155$-$304 is detected with durations of 56 days (Flip method) and 28 days (Baseline method), with both intervals overlapping. This suggests that candidate spectral hardening can occur over a wide range of timescales, from under a month to several hundred days, although the limited number of events prevents any statistical interpretation.

Both blazars have been detected by Imaging Atmospheric Cherenkov Telescopes (IACTs): J0238.4$-$3116 \citep{gate17} and PKS 2155$-$304 \citep{PKS2155}. The latter has been extensively studied in searches for quasi-periodic oscillations \citep[QPOs;][]{penil20,rueda22,penil2025search,rico25}, making it a particularly compelling target for investigating a possible link between spectral hardening and QPOs. Such a connection could help characterize periodic modulations in the jet’s particle acceleration and cooling processes.

\begin{figure*}[h!]

    \includegraphics[width=9cm,height=3.9cm]{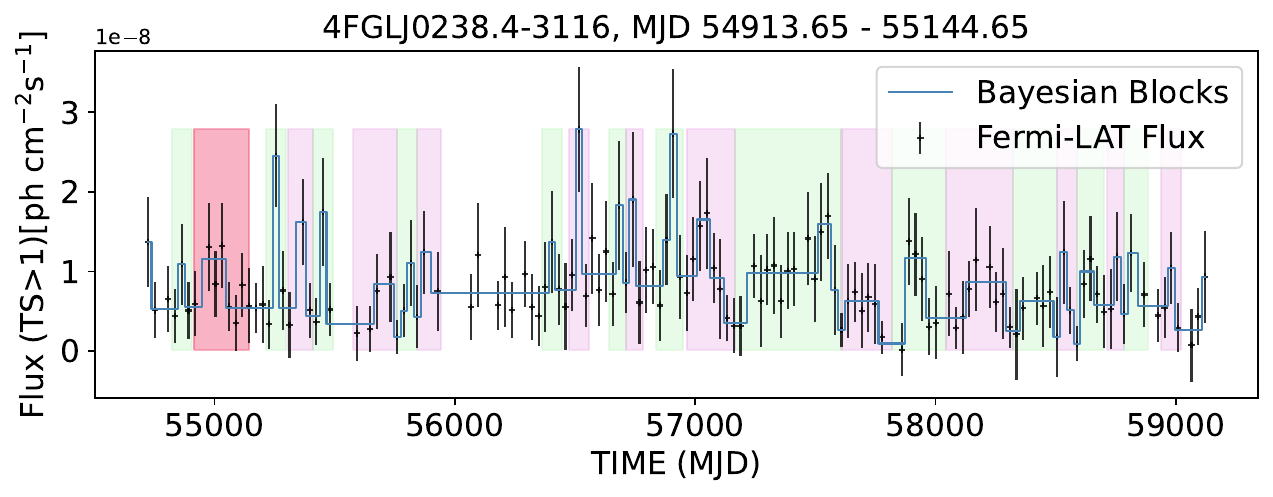}
    \includegraphics[width=9cm,height=3.9cm]{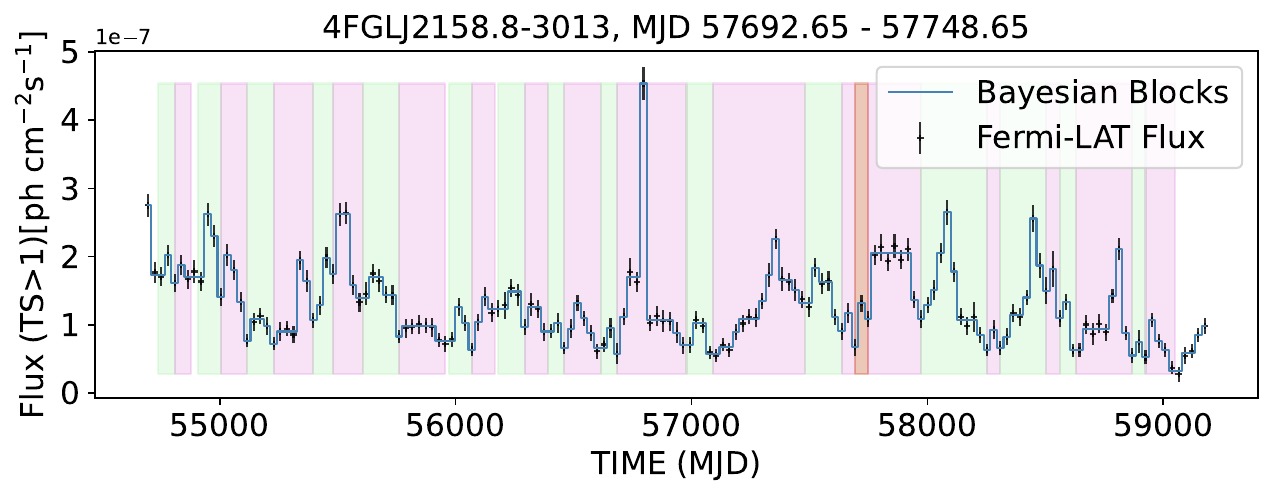}\\
    \includegraphics[width=9cm,height=3.9cm]{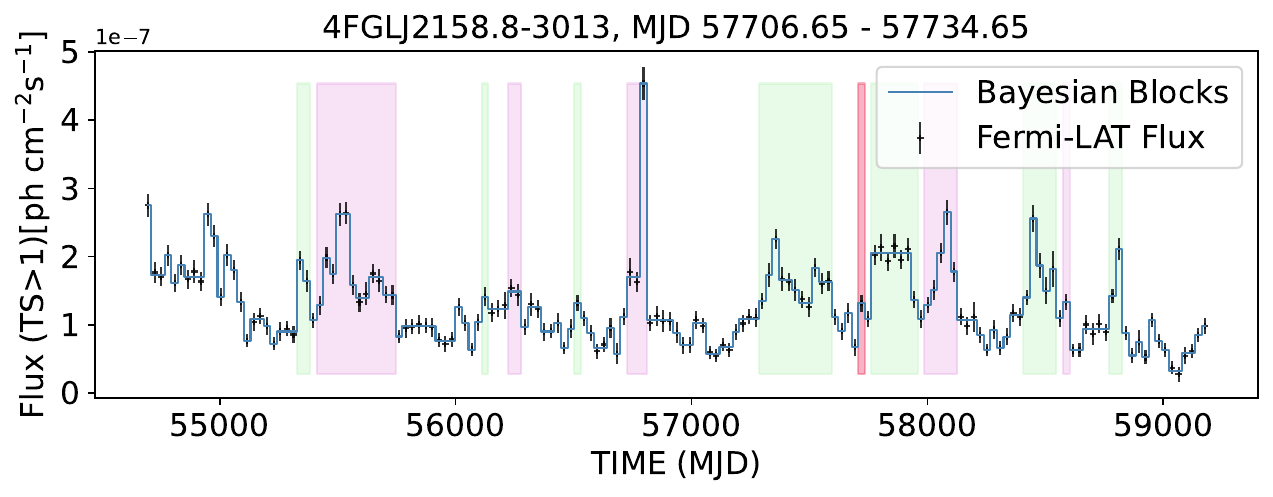}
    
\caption{Light curves of the sources where a spectral hardening feature is detected with TS$_{\mathrm{hardening}} \geq 12$. Insets show the Bayesian Blocks segmentation (blue line), with identified flares shaded alternately in purple and green for clarity. Flares showing spectral hardening are highlighted in red, with the 4FGL source name and corresponding time range labeled above each inset.}\label{figA1}    
\end{figure*}

\begin{figure}
    \centering
    \includegraphics[width=\linewidth]{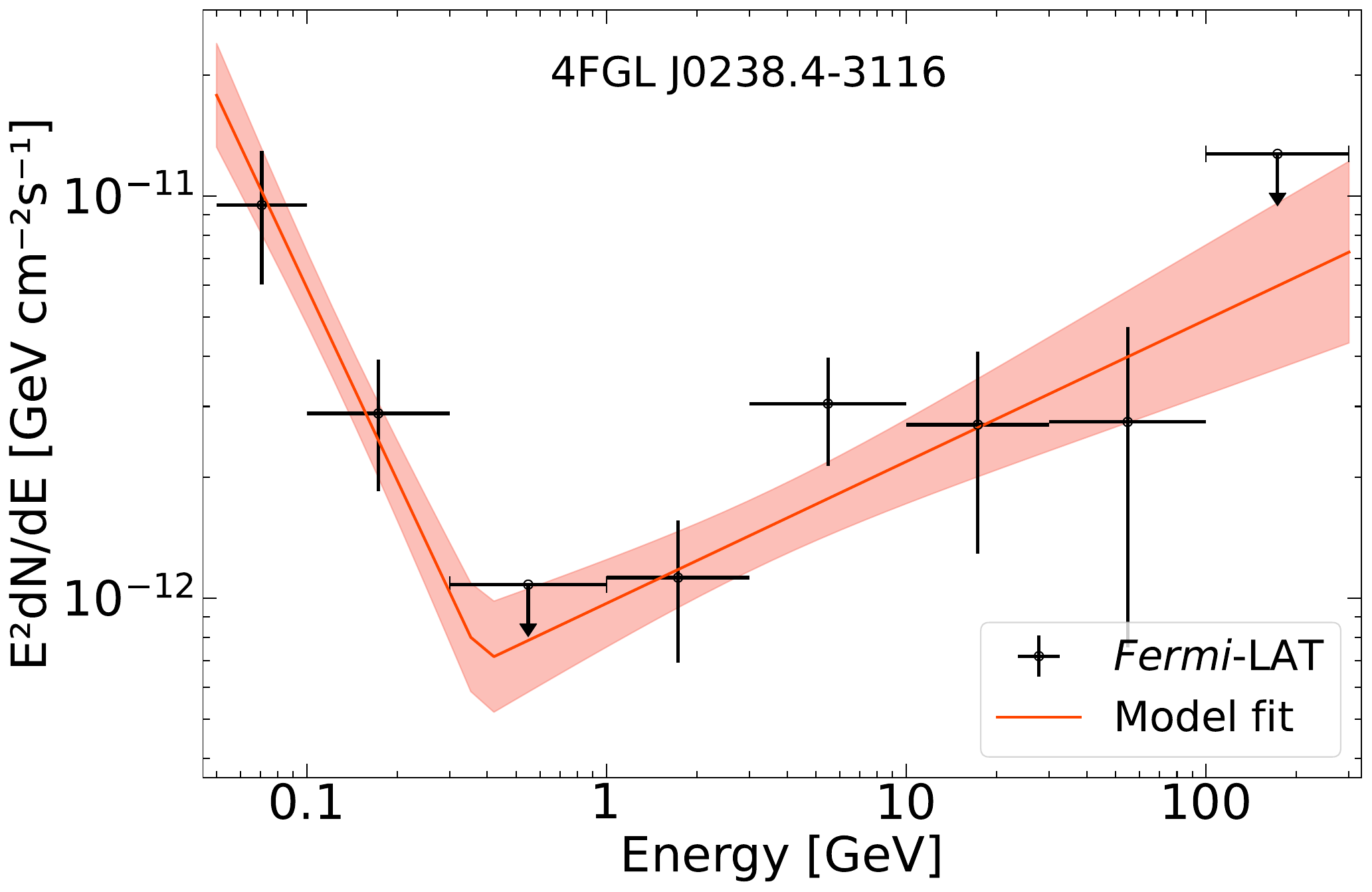}
    
    \vspace{0.1cm}
    \includegraphics[width=\linewidth]{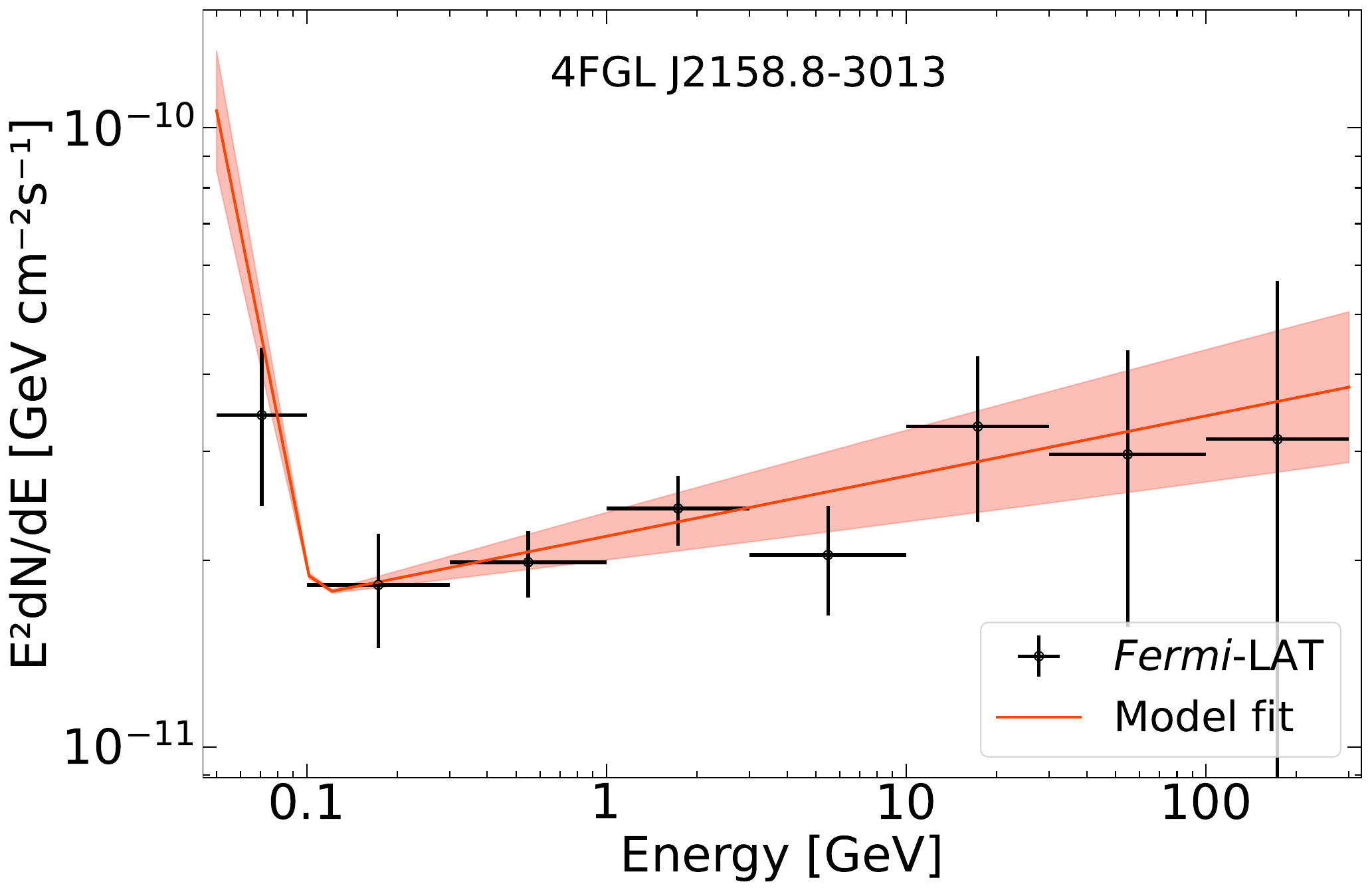}
    
    \vspace{0.1cm}
    \includegraphics[width=\linewidth]{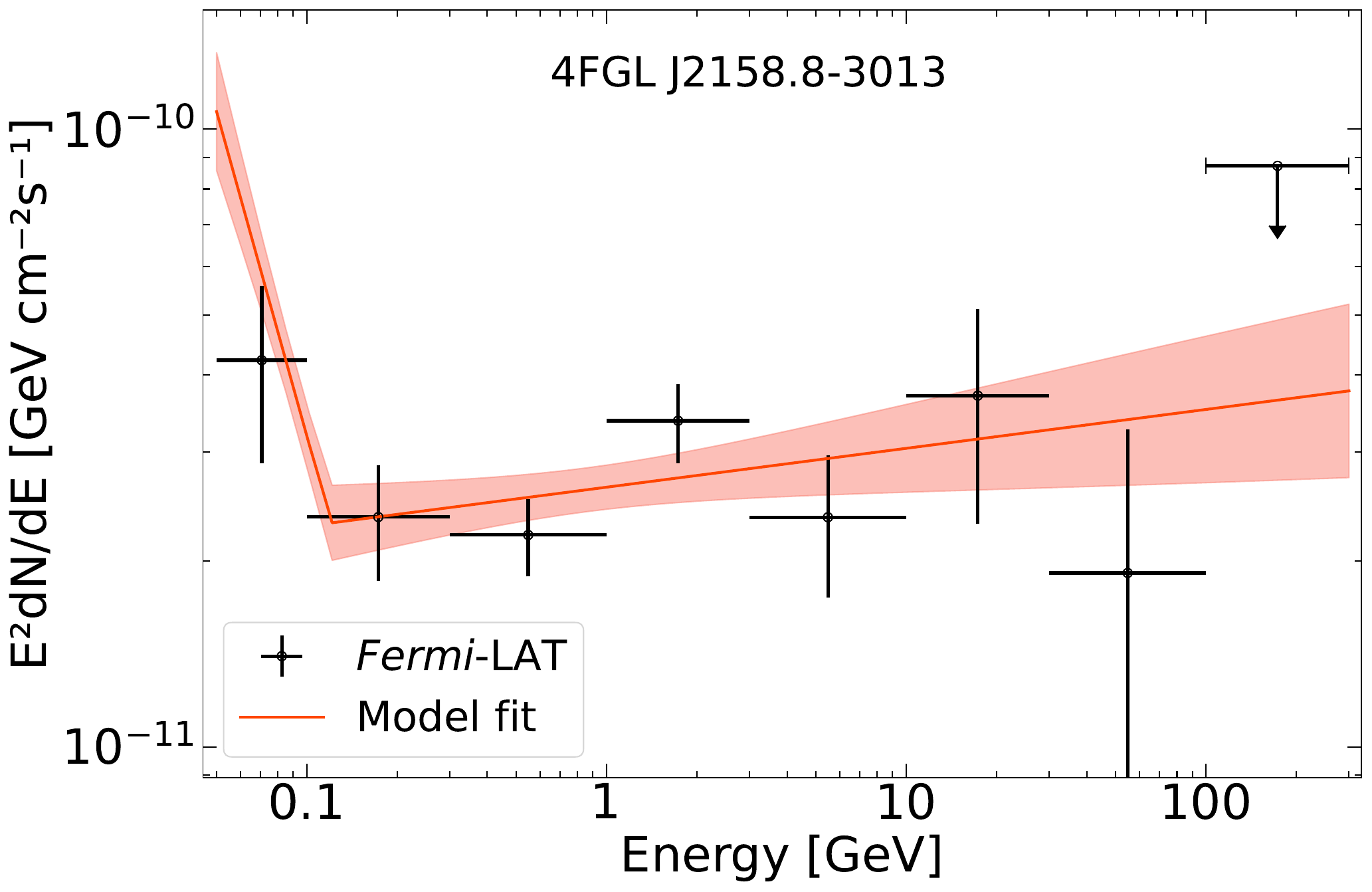}

    \caption{Gamma-ray spectra for the flaring states with TS$_{\mathrm{hardening}} \geq 12$. The {\it Fermi}-LAT data (black dots), with the best-fit BPL model and its uncertainties (red). The PKS 2155$-$304 flare identified by the Flip method (top right) partially overlaps with that selected by the Baseline method (lower left) but spans twice the duration, resulting in a more significant SED for the Flip detection. Low-energy data are more uncertain, though we minimized systematics by using PSF event types and by identifying spectral hardening through likelihood ratios rather than individual flux bins.
    } 
    \label{figA2}
\end{figure}

The LCs, including the selected flaring periods, are displayed in Figure~\ref{figA1}, and the corresponding $\gamma$-ray spectra are shown in Figure~\ref{figA2}. These spectra are provided for illustration only, as the analysis relies exclusively on TS$_{\mathrm{hardening}}$, comparing the BPL and PL likelihoods in LAT data, as described in Section~\ref{sec2.3}.

\subsection{Physical interpretation}
The candidate spectral hardening features identified in this work could, in principle, arise from a limited set of physical processes. In the context of leptonic models, one possibility is a temporary shift of the synchrotron and IC peaks to higher energies, as occasionally suggested during flaring episodes \citep[e.g.,][] {tavecchio2009hard,  zeng2022spectral}. Alternatively, a multi-zone scenario may apply, where a secondary SSC component or compact emission region briefly dominates the spectrum \citep[e.g.,][]{abramowski2012multiwavelength,  abe2024insights}. A third possibility is that magnetic reconnection or other mechanisms inject freshly accelerated electrons, producing unusually hard particle distributions \citep[e.g.,][]{2502.07940v1}. While our search among HSP and EHSP blazars finds, at most, very rare and marginal indications of spectral hardening, it is noteworthy that a recent detection of TeV emission from the FSRQ S5 1027+74 revealed an unusually hard $\gamma$-ray spectrum at GeV, illustrating that such extreme features can occasionally arise even outside the HSP population \citep{paliya2025detection}.

Hadronic or hybrid scenarios have also been invoked in the literature, where proton synchrotron or photohadronic cascades contribute to $\gamma$-ray emission \citep[e.g.,][]{biteau2020progress,2502.07940v1}. However, these generally require extreme physical conditions and remain challenging to test with the current data.

Our analysis alone cannot disentangle between these possibilities. The most robust outcome of our study is that such spectral hardening features are extremely rare, with an occurrence rate below 0.1\% across thousands of flares. This rarity strongly suggests that the physical conditions required to produce sharp breaks in the GeV band are seldom realized in blazar jets. Establishing which of the above scenarios may apply will require future events detected with higher significance and, crucially, simultaneous multi-wavelength observations.

Our pipeline analyzed 4063 flares and identified two independent flaring episodes with indications of spectral hardening: one in 4FGL J0238.4$-$3116 and another in PKS 2155$-$304. The latter is detected with both the Flip and Baseline methods but corresponds to the same flaring activity period. Since both events lie at or just above the detection threshold (TS$_{\mathrm{hardening}} \geq 12$), the occurrence rate of candidate spectral hardening events is below 0.1\%. To our knowledge, there are no theoretical works that provide explicit predictions for the occurrence rate of spectral hardening flares in blazars, as most studies focus on modeling individual events under assumed physical conditions rather than estimating their population frequency \citep[e.g.,~][]{tavecchio2009hard,bottcher2013leptonic}. This confirms that such features are extremely rare in $\gamma$-ray blazars, if present at all. In the context of standard leptonic models, this rarity suggests that the physical conditions required to produce sharp spectral hardening, such as transitions between distinct electron populations or emission zones, are seldom met. Alternatively, the hardening may emerge only under exceptional jet environments involving magnetic reconnection, multi-zone structures, or hadronic contributions. The scarcity of events reinforces the notion that the dominant blazar emission mechanism is well described by smoothly varying power-law spectra across the {\it Fermi}-LAT range, with sharp hardenings representing rare deviations from this behavior.

While these scenarios remain as viable theoretical interpretations, the small number and marginal significance of our candidate events limit any firm conclusions. A definitive physical understanding of spectral hardening will require future detections with high significance and simultaneous multi-wavelength coverage. Coordinated, long-term campaigns by current and upcoming facilities, including H.E.S.S., MAGIC, VERITAS, and CTAO at VHE, will be essential to capture and characterize such extreme events and constrain their underlying mechanisms.

\begin{table*}
\caption{Blazars with TS$_{\mathrm{hardening}}\geq 12$}
\begin{tabular}{llcccccccc}
 \hline
 4FGL Source &  Association & $\nu^{peak}_{s}$ & Period (MJD)  & Method & E$_{b}$ (GeV) & $ \lvert \Delta \Gamma \rvert $ &TS$_{\mathrm{hard}}$\\
Name (1) & (2) & (3) & (4) & (5) & (6) & (7) & (8)\\ 
 \hline
  
 J0238.4$-$3116 & 1RXS J023832.6-311658 & 16.8 & 54913.65-55144.65 & Flip & 0.39$\pm^{0.24}_{0.11}$ & 1.94$\pm{0.32}$ & 12.0 \\
 J2158.8$-$3013 & PKS 2155-304 & 16.2 & 57692.65-57748.65 & Flip & 0.11$\pm^{0.02}_{0.01}$ & 2.54$\pm{0.32}$ & 16.0 \\
 J2158.8$-$3013 & PKS 2155-304 & 16.2 & 57706.65-57734.65 & BL & 0.12$\pm^{0.02}_{0.01}$ & 1.80$\pm{0.34}$ & 12.0 \\
   \hline

\end{tabular}
\tablefoot{A list of 3 events in 2 blazars with TS$_{\mathrm{hardening}}\geq 12$, identified using the Flip and Baseline methods. The two events in PKS 2155$-$304 correspond to the same underlying flare but are detected independently by the two flare selection methods. Therefore, they represent a single flaring episode captured with slightly different boundaries. (1) 4FGL source name, (2) Association with known sources, (3) Synchrotron peak ($\nu_{s}^{peak}$) in log units of Hz, see main text for references, (4) Time period of the event, (5) Method which identified the flare: Flip or Baseline method, (6) break energy in GeV, (7) Difference of the BPL indexes, note that these are negative values, but we list the absolute value and (8) TS$_{\mathrm{hardening}}$.} 

\label{tab1}

\end{table*}

\section{Summary and conclusions}\label{sec5}
In this study, we conducted a systematic search for extreme $\gamma$-ray blazars, focusing on identifying a distinctive spectral hardening, where the $\gamma$-ray spectrum becomes harder at higher energies, with an energy break between 0.1 GeV and 10 GeV, using \textit{Fermi}-LAT data.

We analyzed the LCs of selected high-latitude sources, characterized by synchrotron peaks exceeding 10$^{16}$ Hz, spanning the first 12 years of the {\it Fermi}-LAT mission. Flares within these LCs were systematically selected using two methods, Baseline and Flip, and the presence of spectral hardening was investigated.

Out of 4063 flares identified, our study found three events in two blazars with spectral hardening features at or just above the detection threshold (TS$_{\mathrm{hardening}} \geq 12$), corresponding approximately to a 3$\sigma$ significance level. This implies an occurrence rate of fewer than 0.1\%, providing the first population-level constraint on the frequency of such events in $\gamma$-ray blazars. The scarcity of detections reinforces the notion that the dominant emission in blazar jets is well described by smoothly varying power-law spectra across the Fermi-LAT energy range, with sharp spectral deviations being extremely rare.

One of these events is found in 4FGL J0238.4$-$3116 and the other two in PKS 2155$-$304, the latter likely corresponding to the same physical flare captured by both selection methods. Given the small number of detections and their moderate significance, these findings are statistically compatible with random fluctuations and do not constitute strong evidence for a widespread occurrence of spectral hardening among blazars. Nonetheless, the identified sources remain of interest for follow-up studies, particularly due to their known HSP frequencies and previous detections at very high energies. Notably, both J0238.4$-$3116 and PKS 2155$-$304 have been detected by IACTs. The latter, as a well-studied QPO candidate, stands out as a key target for investigating a possible connection between spectral hardening and QPOs.

Future multi-wavelength observations are essential to further explore these sources, particularly involving IACTs for VHE $\gamma$-ray studies. Our work lays the groundwork for these efforts by identifying key targets and characterizing their tentative spectral hardening behavior. Such observations will be crucial in refining our understanding of the physical processes at play in specific blazar classes and in clarifying the origins of the observed spectral features.

\begin{acknowledgements}
The authors are thankful to D. Yan for private communication on the 1ES 0502+675 analysis. We thank J. Valverde for her helpful revision of the manuscript and J. Ballet, Filippo D'Ammando, and Miguel {\'A}ngel S{\'a}nchez-Conde for fruitful comments. A. Dinesh and J.L.C. acknowledge the support of MCIN project PID2022-138132NB-C42. A. Dom{\'i}nguez is thankful for the support of the project PID2021-126536OA-I00 funded by MCIN/AEI/10.13039/501100011033. This work was supported by the European Research Council, ERC Starting grant \emph{MessMapp} under contract no. $949555$; and by the German Science Foundation DFG, research grant ‘Relativistic Jets in Active Galaxies’ (FOR 5195, grant No. 443220636).\\

The \textit{Fermi}-LAT Collaboration acknowledges generous ongoing support from a number of agencies and institutes that have supported both the development and the operation of the LAT as well as scientific data analysis.
These include the National Aeronautics and Space Administration and the
Department of Energy in the United States, the Commissariat \`a l'Energie Atomique and the Centre National de la Recherche Scientifique / Institut National de Physique Nucl\'eaire et de Physique des Particules in France, the Agenzia Spaziale Italiana and the Istituto Nazionale di Fisica Nucleare in Italy, the Ministry of Education, Culture, Sports, Science and Technology (MEXT), High Energy Accelerator Research Organization (KEK) and Japan Aerospace Exploration Agency (JAXA) in Japan, and the K.~A.~Wallenberg Foundation, the Swedish Research Council and the Swedish National Space Board in Sweden.

Additional support for science analysis during the operations phase is gratefully acknowledged from the Istituto Nazionale di Astrofisica in Italy and the Centre National d'\'Etudes Spatiales in France. This work was performed in part under DOE Contract DE-AC02-76SF00515.

\end{acknowledgements}
\section*{Data availability}
All $\gamma$-ray data are publicly available at the Fermi Science Support Center, including the source catalog used in this work, at this website \url{https://fermi.gsfc.nasa.gov/ssc/}.
\nolinenumbers
\bibliographystyle{aa} 
\bibliography{bib.bib}

\end{document}